\documentclass[twocolumn,aps,prl,superscriptaddress,showpacs]{revtex4}

\usepackage{graphicx}
\usepackage{hyphenat} 

\begin{document}

\newcommand{\etal}{\emph{et al.}}
\newcommand{\Li}{{}^{6}\textrm{Li}}
\newcommand{\LiTwoSthreetwo}{{2S_{1/2}}}
\newcommand{\UVtransition}{{2S_{1/2}\rightarrow 3P_{3/2}}}
\newcommand{\REDtransition}{{2S_{1/2}\rightarrow 2P_{3/2}}}
\newcommand{\GammaTwoP}{\Gamma_{2\mathrm{P}}}
\newcommand{\GammaThreeP}{\Gamma_{3\mathrm{P}}}
\newcommand{\Kfourty}{{}^{40}\textrm{K}}
\newcommand{\Kthirtynine}{{}^{39}\textrm{K}}
\newcommand{\kB}{k_{\mathrm{B}}}
\newcommand{\Limp}{\textrm{Li}\left|3/2,3/2\right\rangle}
\newcommand{\LiOneOne}{\textrm{Li}\left|1/2,1/2\right\rangle}
\newcommand{\IsatTwoP}{I_{\mathrm{sat}}^{2\mathrm{P}}}
\newcommand{\IsatThreeP}{I_{\mathrm{sat}}^{3\mathrm{P}}} 


\title{Two-stage magneto-optical trapping and narrow-line cooling of $\Li$ atoms to high phase\hyp{}space density}

\author{J.~Sebastian}
\affiliation{Centre for Quantum Technologies (CQT), 3 Science Drive 2, Singapore 117543}
\author{Ch.~Gross}
\affiliation{Centre for Quantum Technologies (CQT), 3 Science Drive 2, Singapore 117543}
\author{Ke~Li}
\affiliation{Centre for Quantum Technologies (CQT), 3 Science Drive 2, Singapore 117543}
\author{H.~C.~J.~Gan}
\affiliation{Centre for Quantum Technologies (CQT), 3 Science Drive 2, Singapore 117543}
\author{Wenhui~Li}
\affiliation{Centre for Quantum Technologies (CQT), 3 Science Drive 2, Singapore 117543}
\affiliation{Department of Physics, National University of Singapore, 2 Science Drive 3, Singapore 117542}
\author{K.~Dieckmann}
\email[Electronic address:]{phydk@nus.edu.sg}
\affiliation{Centre for Quantum Technologies (CQT), 3 Science Drive 2, Singapore 117543}
\affiliation{Department of Physics, National University of Singapore, 2 Science Drive 3, Singapore 117542}

\date{\today}

\begin{abstract}
We report an experimental study of peak and phase-space density of a two-stage magneto-optical trap (MOT) of $\Li$ atoms, which exploits the narrower $\UVtransition$ ultra-violet (UV) transition at $323\,$nm following trapping and cooling on the more common D2 transition at $671\,$nm. The UV MOT is loaded from a red MOT and is compressed to give a high phase-space density up to $3\times 10^{-4}$. Temperatures as low as $33\,\mu$K are achieved on the UV transition. We study the density limiting factors and in particular find a value for the light-assisted collisional loss coefficient of $1.3 \pm0.4\times10^{-10}\,\textrm{cm}^3/\textrm{s}$ for low repumping intensity.
\end{abstract}



\maketitle

\begin{figure}
\includegraphics[width=8.0cm]{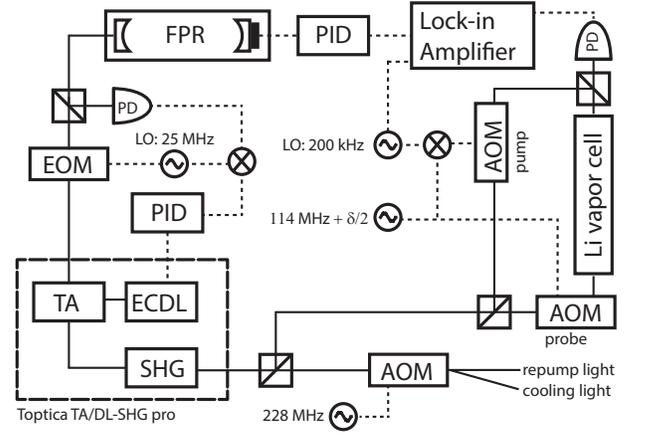}%
\caption{\label{fig:ExperimentalSetup} Frequency stabilized laser system of $323\,$nm. The laser source is a commercial frequency doubling system from Toptica (Toptica TA/DL-SHG pro). From the intracavity frequency doubling stage routinely $35\,$mW of UV light are obtained. The fundamental light at $646\,$nm is produced by an external cavity diode laser (ECDL) and  amplified by a tapered amplifier (TA). This light is stabilized to a high finesse Fabry-Perot resonator (FPR) by means of a Pound-Drever-Hall setup providing fast feedback to the ECDL. The FPR is locked to a lithium vapor cell for long-term stability employing the modulation transfer spectroscopy on the UV transition. The light for the spectroscopy is split into the pump and the probe, which are sent into the spectroscopy cell via two separate AOMs for setting the frequency detuning of the UV light. The error signal is obtained by a lock-in amplifier and used to regulate the piezo element of the FPR.}
\end{figure}

\begin{figure}
\includegraphics[width=8.0cm]{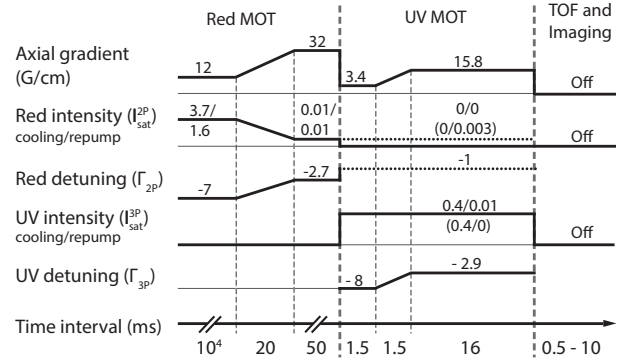}%
\caption{\label{fig:ExperimentalSequence} Experimental sequence and corresponding parameters for red MOT, UV MOT, and imaging as explained in the text. In specifying the red and UV frequency detunings the values apply to both cooling and trapping transitions. The dotted lines for the red intensity and detuning indicate the realization of the UV MOT with the red repumping light, for which the parameters are given in parenthesis.}
\end{figure}

\begin{figure}
 \includegraphics[width=8.0cm]{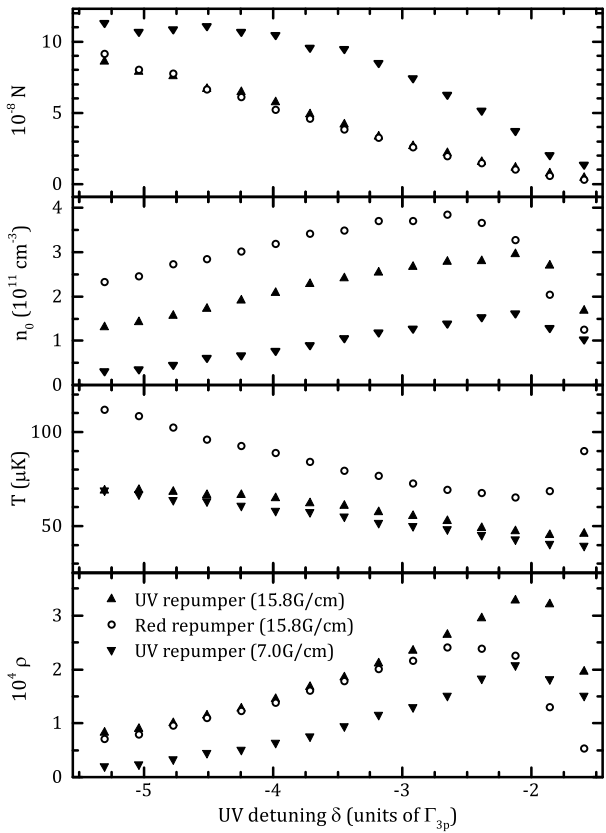}%
 \caption{\label{fig:DensityVsGradient} Performance of the UV MOT. Atom number, peak density, temperature, and phase\hyp{}space density are shown as functions of the detuning with respect to the UV cooling transition. The up/down triangles correspond to the UV MOT with UV repumper at magnetic field gradients of $15.8\,\textrm{G}/\textrm{cm}$ and $7\,\textrm{G}/\textrm{cm}$, respectively. The open circles represent the UV MOT with red repumper at a gradient of $15.8\,\textrm{G}/\textrm{cm}$.}
\end{figure}

\begin{figure}
 \includegraphics[width=8.0cm]{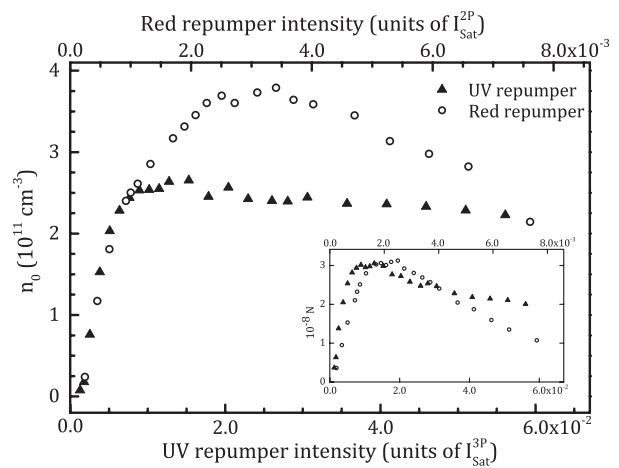}%
 \caption{\label{fig:DensityVsRepupmPower} Peak densities at different intensities for the UV repumper (triangles, lower x\hyp{}axis) and the red repumper (open circles, upper x\hyp{}axis) at a constant UV cooling power. The inset shows the corresponding atom number.}
 \end{figure}

\begin{figure}
\includegraphics[width=8.0cm]{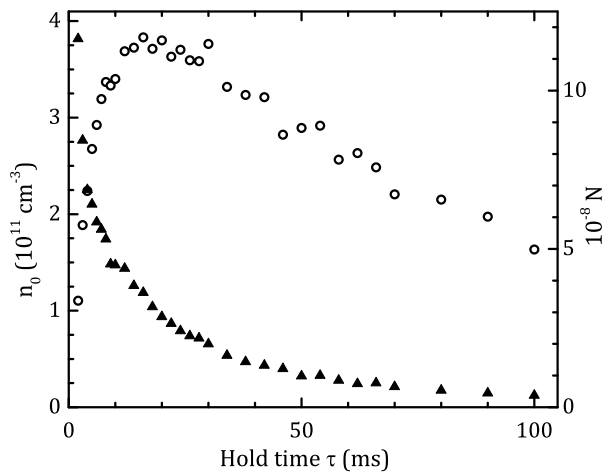}%
\caption{\label{fig:HoldTime} Transient dynamics of the UV MOT after compression. Peak density (circles) and atom number (triangles) are shown versus the hold time in the UV MOT with the final parameters as indicated in Fig.\ref{fig:ExperimentalSequence}.}
\end{figure}

Alkali-metal atoms have been the prime atomic species for studies of degenerate quantum gases due to their simple level structure, well-known atomic properties, and the availability of efficient sub-Doppler laser cooling for most of these elements. While the standard magneto-optical trapping and cooling of alkali-metal atoms is routinely made employing the D2  transition, further laser cooling steps are of great interest particularly for lithium, where a less well-separated hyperfine structure in the optically excited state prevents laser cooling from reaching below the Doppler temperature. Several previously developed schemes have recently been revisited to improve the laser cooling of lithium as well as potassium isotopes to Sub-Doppler temperatures and to make all-optical production of quantum gases more efficient. These schemes utilize gray molasses for $^{39}$K, $^{40}$K, $^{7}$Li, and $^6$Li \cite{Fernandes2012a,Salomon2013,Grier2013,Burchianti2014}, far-detuned light for $^{7}$Li \cite{Hamilton2014}, and narrow transitions to higher optically excited states for $\Li$, $^{40}$K, and $^{41}$K \cite{Duarte2011b,McKay2011a}.\\
Laser cooling on narrow optical transitions was first developed for magneto-optical traps (MOTs) of alkaline-earth atoms using spin-forbidden transitions \cite{Katori1999}. In the case of alkali atoms, the employed narrow transitions are from the ground states $nS_{1/2}$ to the higher excited states $(n+1)P_{3/2}$, which have much smaller dipole matrix elements compared to that of D2 lines. For $\Li$ atoms (see Ref.\cite{Duarte2011b} for the detailed level structure), the $\UVtransition$ transition at the ultra-violet (UV) wavelength of $323\,$nm has an overall linewidth of $\GammaThreeP/\left(2\pi\right)=754\,$kHz, accounting for all decay channels. This corresponds to a Doppler temperature of $18\,\mu$K in contrast to the Doppler temperature of $141\,\mu$K for the $\REDtransition$transition at the red wavelength of $671\,$nm, which has a linewidth of $\GammaTwoP/\left(2\pi\right)=5.9\,$MHz.\\
In this paper, we present our measurements on narrow-line trapping and cooling of fermionic $\Li$ atoms in a two-stage MOT. In particular, we study in detail the loading, cooling and compression of a MOT on the UV transition (UV MOT) following a standard MOT on the D2 transition (red MOT). We optimize the trapping scheme to obtain high phase-space densities. Moreover, we investigate quantitatively the limiting factors for density and atom number such as photon-reabsorption, light assisted collisional loss, and photo-ionization.\\
A standard diode laser system is used to produce the six beams for the red MOT, each of which has a $1/e^2$ radius of $9\,$mm and contains both cooling and repumping frequencies separated by the ground state hyperfine splitting of $228\,$MHz. In the following we refer to the light driving the $2S_{1/2},F=1/2\rightarrow 2P_{3/2}$ transition as the red repumper. In order to take advantage of the narrow-line to achieve low MOT temperatures, we produce $323\,$nm light with a linewidth below that of the UV transition. The UV laser system, as illustrated in Fig.\ref{fig:ExperimentalSetup}, is based on a commercial frequency doubling system. The linewidth reduction is achieved by locking the fundamental light of the laser system to a high-finesse Fabry-Perot resonator (FPR) with the Pound-Drever-Hall method. The FPR finesse of $420$ yields a linewidth of $70\,$kHz on short timescales for the UV light. For the long-time frequency stabilization, the FPR is locked to a spectroscopy cell using the UV transition. The main output of the laser passes through an acousto-optic modulator (AOM) at $228\,$MHz to generate the cooling and the repumping light for the UV MOT. The latter drives the $2S_{1/2},F=1/2\rightarrow 3P_{3/2}$ transition and will be referred to as the UV repumper. The frequencies of both beams are simultaneously tuned together by AOMs in the spectroscopy setup. The two UV beams are combined on a polarizing beamsplitter and expanded to have a $1/e^2$ radius of $4.3\,$mm. This combined UV laser beam is then split into three beams, which are overlapped with three red MOT beams on dichroic mirrors and retro-reflected after passing through the atomic cloud.\\
The experimental sequence, illustrated in Fig.\ref{fig:ExperimentalSequence}, comprises of loading and compression in a red MOT, transferring into a UV MOT, and further cooling and compression in the UV MOT.\\
For capturing atoms from a Zeeman\hyp{}slowed beam into the red MOT we use six beams, each with a peak intensity of $3.7\,\IsatTwoP$ for the cooling and $1.6\,\IsatTwoP$ for the repumping transitions. The saturation intensity for the red transition is $ \IsatTwoP = 2.54\,\textrm{mW}/\textrm{cm}^2 $. During the loading phase the detuning $\delta$ is $ - 7 \, {\GammaTwoP} $ and an axial magnetic field gradient of $12\,\textrm{G} / \textrm{cm}$ is applied. After $10 \, \textrm{s}$ of loading the atomic cloud contains around $ 3 \times 10^{9}$  atoms at a temperature of $ 1.4 \, \textrm{mK} $, which is much higher than the Doppler temperature. To facilitate loading into the small volumed UV MOT, a compression of the red MOT (CMOT) is carried out by simultaneously changing the detuning to $-2.7\, \GammaTwoP$, reducing the intensity to around $ 0.01\, \IsatTwoP $, and ramping the magnetic field gradient to $32\,\textrm{G} / \textrm{cm} $. After this CMOT stage the temperature of the atomic cloud is reduced to $ 274\,\textrm{$\mu$K}$ and $1.8 \times 10^9 $  atoms remain trapped. We determine a peak density $n_0=6.4 \times 10^{10}\, \textrm{cm}^{-3} $ and a phase\hyp{}space density $ \rho=n_0 h^3 (2\pi m k_B T)^{-3/2} = 5.1 \times 10^{-6}$.\\
The atoms are then transferred to the UV MOT by switching on the UV beams and rapidly reducing the magnetic field gradient while the red MOT light is extinguished. Each of the UV beams has a peak intensity of $0.4\,\IsatThreeP$ for the cooling and $0.01\,\IsatThreeP$ for the repumping transitions, where the saturation intensity of the UV transition is $\IsatThreeP = 13.8\,\textrm{mW}/\textrm{cm}^2$. We achieve the highest transfer efficiency with a reduced magnetic field gradient of $3.4\,\textrm{G}/\textrm{cm} $ and a UV light detuning of $-8\,\GammaThreeP$. At these parameters the lifetime of the UV MOT is short, and $1.5\,\textrm{ms}$ after the transfer we implement another CMOT stage. For this the gradient and the detuning are linearly ramped to the final values within $1.5\,\textrm{ms}$, while keeping the intensities constant. After a hold time $\tau$ of $16\,\textrm{ms}$, allowing for the spatial compression of the atomic cloud to be completed, the atoms are probed by standard absorption imaging on the D2 line to infer atom number, density, and temperature. Instead of using the UV repumper we also investigate the UV MOT by using the red repumper with a constant detuning of $ -1\, \GammaTwoP$ throughout the UV MOT stage.\\
The experimental parameters for the final stage of the UV CMOT are optimized by scanning the detuning of the UV laser for various final magnetic field gradients. Peak density and temperature are measured by probing the atomic cloud at $ 0.5 \, \textrm{ms}$ and $ 8 \, \textrm{ms}$ time-of-flight, respectively. These values together with atom number and phase-space density are plotted in Fig.\ref{fig:DensityVsGradient}. A set of plots with the red repumper replacing the UV repumper are also shown. Different magnetic field gradients do not seem to give significantly different temperatures. The highest peak density of $3\times 10^{11}\,\textrm{cm}^{-3} $ is observed at a magnetic field gradient of $15.8\,\textrm{G}/\textrm{cm}$. Lower gradients lead to lower peak densities, as exemplarily shown for a gradient of $7\,\textrm{G}/\textrm{cm}$. However, increasing the magnetic field gradient beyond $15.8\,\textrm{G}/\textrm{cm}$ does not much further increase the peak density, as lower atom numbers are attained. Even higher peak densities are achieved by using the red repumper, for which the temperature is comparatively high and hence the phase-space density is reduced. The highest phase\hyp{}space density of the UV MOT is attained with the UV repumper at a temperature of $47\,\mu\textrm{K}$ and an atom number of $1\times 10^{8}$. The resulting value of $ \rho= 3.3 \times 10^{-4}$ is an order of magnitude higher as compared to previous measurements \cite{Duarte2011b,Burchianti2014} for $\Li$ atoms. We also optimized the parameters to obtain temperatures as low as $33\,\mu$K by reducing the UV intensity to $0.03\,\IsatThreeP$ and by lowering the gradient to $2.1\,\textrm{G}/\textrm{cm}$.\\  
We find that, for either the UV or red repumper, the peak density depends strongly on the repumper power. We study this effect at a gradient of $15.8\,\textrm{G}/\textrm{cm}$ and a UV detuning of $-2.9\,\GammaThreeP$. While the cooling power is kept constant, the peak density is measured as a function of the repumper intensity, as shown in Fig.\ref{fig:DensityVsRepupmPower}. As the hyperfine structure of the optically excited states is not resolved for lithium, the repumper intensity needs to be sufficiently strong. At very low intensities the trapping force is decreased due to the reduced scattering rate on the UV transition, and atoms are lost from the trap as can be seen in the inset of Fig.\ref{fig:DensityVsRepupmPower}.\\
If we use the red repumper, we obtain the highest peak density of $3.8\,\times 10^{11}\,\textrm{cm}^{-3}$ at an intensity of $ \simeq 0.003\, \IsatTwoP $, corresponding to only $10\,\mu\textrm{W}$ per beam. At larger repumper intensity a large fraction of the atoms is optically pumped out of the $ F=1/2$ ground state. Hence, the total scattering rate on the UV transition is increased and along with it density limiting factors like photon\hyp{}reabsorption and light\hyp{}assisted collisional two\hyp{}body losses are enhanced. Obtaining high density at comparatively low repumper intensity therefore resembles the situation of a dark MOT.\\
For the UV MOT with UV repumper, after reaching a maximum peak density of $2.5\times 10^{11}\,\textrm{cm}^{-3} $ at $ \simeq 0.009\, \IsatThreeP$, corresponding to a ratio of $44:1$ between cooling and repumper intensity, the density remains approximately constant. One reason for this different trend could be the overall lower observed peak density for the UV MOT with UV repumper, which leads to lower two\hyp{}body losses and correspondingly a slower decay in the atom number.\\
It is important to note that the observed high peak density is a transient effect. Shown in Fig.\ref{fig:HoldTime} are measurements of peak density and atom number as functions of hold time $\tau$ after gradient and detuning reached the final values. In the initial phase we observe a fast contraction of the atomic cloud, resulting in a maximal peak density after approximately $16\,\textrm{ms}$, which then decays at a slower rate. This compression timescale is expected, if a model of overdamped oscillatory motion is assumed for a single atom in the light field. This density buildup is limited by photon\hyp{}reabsorption and two\hyp{}body losses, which first lead to a reduction in the atom number and eventually also in the peak density, similar as described in \cite{DePue2000}. The observed peak densities are therefore difficult to be increased by a higher atom number. We verified this by changing the loading time. For the case of the red repumper the atom number can be lowered by a factor of $2$, before the peak density is noticeably reduced.\\
In order to further study the limitations on density and atom number we perform lifetime measurements of the compressed UV MOT with either the UV or red repumper to determine the one\hyp{} and two\hyp{}body loss coefficients. This is done at repumping powers leading to high densities and at a gradient of $15.8\,\textrm{G}/\textrm{cm}$ and detuning of $-2.9\,\GammaThreeP$. The time evolution of the number of atoms $N$ in an atomic cloud with a Gaussian density distribution is described by \cite{Weiner1999}
\begin{equation}
\label{eq:LossRateGaussian}
\frac{dN}{dt}=-\gamma N-\beta \frac{N^2}{\left( 2\pi \right)^{3/2}r_{\rho}^2 r_{z}}\,,
\end{equation}
where $r_{\rho}$ ($r_{z}$) is the radial (axial) $1/e$ radius of the density distribution and $\gamma$ and $\beta$ are the one\hyp{} and two\hyp{}body loss coefficients. We observe that the radius of the cloud is not constant in the atom number, leading to a time dependence of the radius $r_{i} = r_{i}\left(t\right)$, $\left(i=\rho, z \right)$. This implies that Eq.~(\ref{eq:LossRateGaussian}) cannot be solved analytically. Instead, the differential equation is fitted directly to the experimental data, in order to extract the loss coefficients. We obtain $\beta=1.3 \pm0.4\times10^{-10}\,\textrm{cm}^3/\textrm{s}$ if the UV repumper is used and $\beta=2.1\pm0.6\times10^{-10}\, \textrm{cm}^3/\textrm{s}$ for the red repumper. These values are high compared to previous measurements for the red MOT \cite{Ridinger2011}, but comparable to values reported for a potassium MOT driven on the narrow-line transition \cite{McKay2011a}. We attribute the larger two\hyp{}body loss rate to the reduced effective trap depth \cite{Ritchie1995} and the small capture velocity, due to the low intensity of the beams and narrow linewidth of the UV transition.\\
We obtain high one\hyp{}body decay rates $\gamma$ of $1/\left(84\,\textrm{ms}\right)$ and $1/\left(224\,\textrm{ms}\right)$ for using the UV and red repumper, respectively. We can exclude losses due to collisions with background atoms as a dominant contribution, as decay measurements of the red MOT yield lifetimes on the order of $30\,\textrm{s}$. We attribute the short lifetime to the finite capture velocity for the parameters after the compression ramp, which are optimized to achieve high phase-space density. This is consistent with our observation of longer one\hyp{}body decay times, if the gradient is reduced and the detuning increased. Driving the MOT on the UV transition introduces decay channels via photoionization. The loss rate due to ionization is estimated based on a theoretical calculation of the ionization cross section of the excited states \cite{Lahiri1993}. We find approximately $1\,$s as a lower limit for the lifetime with respect to photoionization and conclude that this process is negligible on the timescale of the UV MOT and future loading into a dipole trap.\\
In summary we have implemented a two-stage $\Li$ MOT utilizing the narrow UV transition to achieve sub-Doppler temperatures  with respect to the D2 transition. By compression in the UV MOT we achieve a transient phase of high density. This is possible as the rapid decay of the atom number is initially compensated by the contraction of the atomic cloud. We have measured high collisional two\hyp{}body loss rates, which we identify as an important density limiting factor. Our optimization results in the highest phase-space density for laser cooled lithium atoms reported so far. This improves the efficiency of direct loading from the MOT into an optical dipole trap and facilitates further cooling for the production of a quantum degenerate sample.\\
We gratefully acknowledge funding by the Singapore National Research Foundation and the Ministry of Education.


\begin{thebibliography}{10}

\bibitem{Fernandes2012a}
D.~Rio Fernandes, F. Sievers, N. Kretzschmar, S. Wu, C. Salomon and F. Chevy, Europhys. Lett. {\bf 100}, 63001 (2012).

\bibitem{Grier2013}
A.~T Grier, I. Ferrier-Barbut, B.~S. Rem, M. Delehaye, L. Khaykovich, F. Chevy, and C. Salomon, Phys. Rev. A {\bf 87}, 063411 (2013).

\bibitem{Salomon2013}
G. Salomon, L. Fouch\'{e}, P. Wang, A. Aspect, P. Bouyer, and T. Bourdel, Europhys. Lett. {\bf 104}, 63002 (2013).

\bibitem{Burchianti2014}
A. Burchianti, G. Valtolina, J.~A. Seman, E. Pace, M. De Pas, M. Inguscio, M. Zaccanti, G. Roati, arXiv:1406.4788 (2014).

\bibitem{Hamilton2014}
P. Hamilton, G. Kim, T. Joshi, B. Mukherjee, D. Tiarks, and H. M\"uller, Phys. Rev. A {\bf 89}, 023409 (2014).

\bibitem{Duarte2011b}
P.~M. Duarte,  R.~A. Hart, J.~M. Hitchcock, T.~A. Corcovilos, T.-L. Yang, A. Reed, and  R.~G. Hulet, Phys. Rev. A {\bf 84}, 061406 (2011).

\bibitem{McKay2011a}
D.~C. McKay, D. Jervis, D.~J. Fine, J.~W. Simpson-Porco, G.~J.~A. Edge, and J.~H. Thywissen,  Phys. Rev. A {\bf 84}, 063420 (2011).

\bibitem{Katori1999}
H. Katori, T. Ido, Y. Isoya, and M. Kuwata-Gonokami, Phys. Rev. Lett. {\bf 82}, 1116 (1999).

\bibitem{DePue2000}
M.~T. DePue, S.~L. Winoto, D.~J. Han, and D.~S. Weiss, Opt. Commun. {\bf 180}, 73-79 (2000).

\bibitem{Ritchie1995}
N.~W.~M. Ritchie, E.~R.~I. Abraham, Y.~Y. Xiao, C.~C. Bradley, R.~G. Hulet, and P.~S. Julienne, Phys. Rev. A {\bf 51}, R890--R893 (1995).

\bibitem{Ridinger2011}
A. Ridinger, S. Chaudhuri, T. Salez, U. Eismann, D.~R. Fernandes, K. Magalh\~{a}es, D. Wilkowski, C. Salomon, and F. Chevy, Eur. Phys. J. D. {\bf 65}, 223-242 (2011).

\bibitem{Lahiri1993}
J. Lahiri, S.~T. Manson, Phys. Rev. A {\bf 48}, 3674-3679 (1993).

\bibitem{Weiner1999}
J. Weiner, V.~S. Bagnato, S. Zilio, and P.~S. Julienne, Rev. Mod. Phys., {\bf 71}, 1-85 (1999).

\end{thebibliography}
\end{document}